\documentclass[11pt,a4paper,twoside]{article}
\usepackage{graphicx}
\usepackage{color}
\usepackage{hyperref}
\begin{document}
\baselineskip=0.20in
\vspace{20mm}
\baselineskip=0.30in
{\bf \LARGE
\begin{center}
Approximate eigensolutions of the deformed Woods-Saxon potential via AIM
\end{center}}
\vspace{4mm}
\begin{center}
{\Large {\bf Sameer M. Ikhdair}}\footnote{\scriptsize E-mail:~ sikhdair@neu.edu.tr} \\
\end{center}
{\small
\begin{center}
{\it Physics Department, Faculty of Science, An-Najah National University, \\
Nablus, West Bank, Palestine.}\\
and\\
{\it Physics Department, Near East University, 922022 Nicosia, Northern Cyprus, Turkey.}
\end{center}}
\vspace{4mm}
\begin{center}
{\Large {\bf Babatunde J. Falaye}}\footnote{\scriptsize E-mail:~ fbjames11@physicist.net} \\
\small
\vspace{2mm}
{\it Theoretical Physics Section, Department of Physics\\
 University of Ilorin,  P. M. B. 1515, Ilorin, Nigeria.}
\vspace{4mm}
\end{center}
\begin{center}
{\Large {\bf Majid Hamzavi\dag}}\footnote{\scriptsize E-mail:~ majid.hamzavi@gmail.com ( \dag Corresponding author).} \\
\small
\vspace{2mm}
{\it Department of Science and Engineering, Abhar Branch, Islamic Azad University, Abhar, Iran}
\end{center}
\vspace{12mm}
\noindent
\begin{abstract}
\noindent
By using the Pekeris approximation, the Schr\"{o}dinger equation is solved for the nuclear deformed Woods-Saxon potential within the framework of the asymptotic iteration method (AIM).  The energy levels are worked out and the corresponding normalized eigenfunctions are obtained in terms of hypergeometric function. 
\end{abstract}


iteration method.

{\bf PACs No.} 03.65.Pm; 03.65.Ge; 03.65.-w; 03.65.Fd; 02.30.Gp

The deformed Woods-Saxon (dWS) potential is a short range potential and widely used in nuclear, particle, atomic, condensed matter and chemical physics [1-7]. This potential is reasonable for nuclear shell models and used to represent the distribution of nuclear densities. The dWS and spin-orbit interaction are important and applicable to deformed nuclei \cite{BJ8} and to strongly deformed nuclides \cite{BJ9}. The dWS potential parameterization at large deformations for plutonium $^{237, 239, 241}Pu$  odd isotopes was analyzed \cite{BJ10}. The structure of single-particle states in the second minima of $^{237, 239, 241}Pu$ has been calculated with  dWS potential. The Nuclear shape was parameterized. 

The parameterization of the spin-orbit part of the potential was obtained in the region corresponding to large deformations (second minima) depending only on the nuclear surface area. The spin-orbit interaction of a particle in a non-central self consistent field of the WS type potential was investigated for light nuclei and the scheme of single-particle states has been found for mass number $A_0=10$ and $25$ \cite{BJ8}. Two parameters of the spin-orbit part of the dWS potential, namely the strength parameter and radius parameter were adjusted to reproduce the spins for the values of the nuclear deformation parameters \cite{BJ11}.

Badalov et al. investigated Woods-Saxon potential in the framework of Schr\"{o}dinger and Klein-Gordon equations by means of Nikiforov-Uvarov method \cite{BJ12, BJ13}. In our recent works \cite{BJ14, BJ15}, we have studied the relativistic Duffin-Kemmer-Petiau and the Dirac equation for dWS potential. In addition, we have also obtained the bound state solutions of the $PT-/non-PT$-symmetric and non-Hermitian modified Woods-Saxon potential with the real and complex-valued energy levels \cite{BJS1}. 

The motivation of this work is to study the Schr\"{o}dinger equation with dWS potential for any arbitrary orbital quantum number $\ell$. We obtain analytical expressions for the energy levels and wave functions in closed form via asymptotic iteration method (AIM).

We briefly outline the AIM with the details can be found in references \cite{BJ16, BJ17}.

AIM is proposed to solve the homogenous linear second-order differential equation of the form
\begin{equation}
y_n''(x)=\lambda_0(x)y_n'(x)+s_0(x)y_n(x),
\label{E1}
\end{equation}
where $\lambda_0(x)\neq0$ and the prime denotes the derivative with respect to $x$, the extral parameter $n$ is thought as a radial quantum number. The variables, $s_0(x)$ and  $\lambda_0(x)$ are sufficiently differentiable. To find a general solution to this equation, we differentiate equation (\ref{E1}) with respect to $x$ as
\begin{equation}
y_n'''(x)=\lambda_1(x)y_n'(x)+s_1(x)y_n(x),
\label{E2}
\end{equation}
where
\begin{eqnarray}
\lambda_1(x)&=&\lambda_0'(x)+s_0(x)+\lambda_0^2(x),\nonumber\\
s_1(x)&=&s_0'(x)+s_0(x)\lambda_0(x),
\label{E3}
\end{eqnarray}
and the second derivative of equation (\ref{E1}) is obtained as
\begin{equation}
y_n''''(x)=\lambda_2(x)y_n'(x)+s_2(x)y_n(x),
\label{E4}
\end{equation}
where
\begin{eqnarray}		
\lambda_2(x)&=&\lambda_1'(x)+s_1(x)+\lambda_0(x)\lambda_1(x),\nonumber\\
	 s_2(x)&=&s_1'(x)+s_0(x)\lambda_1(x).
\label{E5}
\end{eqnarray}
Equation (\ref{E1}) can be iterated up to $(k+1)th$ and $(k+2)th$ derivatives, $k=1,2,3...$ Therefore we have
\begin{eqnarray}
y_n^{(k+1)}(x)&=&\lambda_{k-1}(x)y_n'(x)+s_{k-1}(x)y_n(x),\nonumber\\
y_n^{(k+2)}(x)&=&\lambda_{k}(x)y_n'(x)+s_{k}(x)y_n(x),
\label{E6}
\end{eqnarray}
where
\begin{eqnarray}
\lambda_k(x)&=&\lambda_{k-1}'(x)+s_{k-1}(x)+\lambda_o(x)\lambda_{k-1}(x),\nonumber\\
s_k(x)&=&s_{k-1}'(x)+s_0(x)\lambda_{k-1}(x).
\label{E7}
\end{eqnarray}
From the ratio of the $(k+2)$th and $(k+1)$th derivatives, we obtain
\begin{equation}
\frac{d}{dx}ln\left[y_n^{k+1}(x)\right]=\frac{y_n^{(k+2)}(x)}{y_n^{(k+1)}(x)}=\frac{\lambda_k(x)\left[y_n'(x)+\frac{s_k(x)}{\lambda_k(x)}y_n(x)\right]}{\lambda_{k-1}(x)\left[y_n'(x)+\frac{s_{k-1}(x)}{\lambda_{k-1}(x)}y_n(x)\right]},
\label{E8}
\end{equation}
if $k>0$, for sufficiently large $k$, we obtain $\alpha$ values from
\begin{equation}
\frac{s_k(x)}{\lambda_k(x)}=\frac{s_{k-1}(x)}{\lambda_{k-1}(x)}=\alpha(x),
\label{E9}
\end{equation}
with quantization condition
\begin{equation}
\delta_k(x)=
\left|
\begin{array}{lr}     
\lambda_k(x)&s_k(x) \\      
  \lambda_{k-1}(x)&s_{k-1}(x)
  \end{array}
  \right|=0\ \ ,\ \  \ k=1, 2, 3....
\label{E10}
  \end{equation}
Then equation (\ref{E8}) reduces to 
\begin{equation}
\frac{d}{dx}ln\left[y_n^{(k+1)}(x)\right]=\frac{\lambda_k(x)}{\lambda_{k-1}(x)},
\label{E11}
\end{equation}
which yields the general solution of Eq. (\ref{E1})
\begin{equation} y_n(x)=\exp\left(-\int^x\alpha(x')dx'\right)\left[C_2+C_1\int^x\exp\left(\int^{x'}\left[\lambda_o(x'')+2\alpha(x'')\right]dx''\right)dx'\right].
\label{E12}
\end{equation}
For a given potential, the idea is to convert the radial Schr$\ddot{o}$dinger equation to the form of equation (\ref{E1}). Then $\lambda_o(x)$ and $s_o(x)$ are determine and $s_k(x)$ and $\lambda_k(x)$ parameters are calculated by the recurrence relations given by equation (\ref{E7}). The energy eigenvalues are then obtained by the condition given by equation (\ref{E10}) if the problem is exactly solvable.

Now, suppose we wish to solve the radial Schr$\ddot{o}$dinger equation for which the homogenous linear second-order differential equation takes the following general form
\begin{equation}
y''(x)=2\left(\frac{tx^{N+1}}{1-bx^{N+2}}-\frac{m+1}{x}\right)y'(x)-\frac{Wx^N}{1-bx^{N+2}},
\label{E13}
\end{equation}
where $t$, $m$ and $W$ are arbitrary constant. The exact solution $y_n(x)$ can be expressed as $\cite{BJ17}$
\begin{equation}
y_n(x)=(-1)^nC_2(N+2)^n(\sigma)_{_n}{_2F_1(-n,\rho+n;\sigma;bx^{N+2})},
\label{E14}
\end{equation}
where the following notations has been used
\begin{equation}
(\sigma)_{_n}=\frac{\Gamma{(\sigma+n)}}{\Gamma{(\sigma)}}\ \ ,\ \ \sigma=\frac{2m+N+3}{N+2}\ \ and\ \ \rho=\frac{(2m+1)b+2t}{(N+2)b}.
\label{E15}
\end{equation}

The deformed Woods-Saxon potential we investigate in this study is defined as \cite{BJ14, BJ15, N1}
\begin{equation}
V(r)=-\frac{V_0}{q+\exp{(\frac{r-R}{a})}},\ \ \ \ R=r_0A_0^{1/3},\ \ \ \ V_0=(40.5+0.13A_0)MeV,\ \ \ \ R>>a,\ \ \ \ q>0,
\label{E16}
\end{equation}
where $V_0$ is the depth of potential, $q$ is a real parameter which determines the shape (deformation) of the potential, $a$ is the diffuseness of the nuclear surface, $R$ is the width of the potential, $A_0$ is the atomic mass number of target nucleus and $r_0$ is radius parameter. By inserting this potential into the Schr$\ddot{o}$dinger equation \cite{N2, BJS2} as
\begin{equation}
\left(-\frac{\hbar^2}{2\mu}\left[\frac{1}{r^2}\frac{\partial}{\partial r}r^2\frac{\partial}{\partial r}+\frac{1}{r^2\sin \theta}\frac{\partial}{\partial\theta}\left(\sin\theta\frac{\partial}{\partial\theta}\right)+\frac{1}{r^2\sin^2\theta}\frac{\partial^2}{\partial\phi^2}\right]+V(r)\right)\Psi_{n\ell m}(r)=E\Psi_{n\ell m}(r),
\label{E17}
\end{equation}
and setting the wave functions $\Psi_{n\ell m}(r)=r^{-1}R_{n\ell}(r)Y_{\ell m}(\theta,\phi)$, we obtain the radial part of the equation by the separation of variables as
\begin{equation}
\left[\frac{d^2}{dr^2}+\frac{2\mu}{\hbar^2}\left(E_{n\ell}+\frac{V_0}{q+\exp{(\frac{r-R}{a})}}\right)-\frac{\ell(\ell+1)}{r^2}\right]R_{n\ell}(r)=0.
\label{E18}
\end{equation}
Because of the total angular momentum centrifugal term, equation (\ref{E18}) cannot be solved analytically for $\ell\neq0$. Therefore, we shall use the Pekeris approximation in order to deal with this centrifugal term and so we may express it as follows $\cite{BJ14, BJ18, BJ19}$
\begin{equation}
U_{cent.}(r)=\frac{1}{r^2}=\frac{1}{R^2\left(1+\frac{x}{R}\right)^2}\cong\frac{1}{R^2}\left(1-2\left(\frac{x}{R}\right)+3\left(\frac{x}{R}\right)^2+\cdots\right),
\label{E19}
\end{equation}
with $x=r-R$. In addition, we may also approximately express it in the following way
\begin{equation}
\tilde{U}=\frac{1}{r^2}\cong\frac{1}{R^2}\left[D_0-\frac{D_1}{q+\exp{(\nu x)}}+\frac{D_2}{\left(q+\exp{(\nu x)}\right)^2}\right],
\label{E20}
\end{equation}
where $\nu=1/a$. After expanding (\ref{E20}) in terms of $x$, $x^2$, $x^3$,$\cdots$ and next, comparing with equation (\ref{E19}), we obtain expansion coefficients $D_0$, $D_1$ and $D_2$ as follows:
\begin{eqnarray}
D_0=1-\left[\frac{1+exp(-aR)}{aR}\right]^2\left[\frac{4aR}{1+exp(-aR)}-3-aR\right],\nonumber\\ 
D_1=2[exp(aR)+1]\left[\frac{3(1+exp(-aR))}{aR}-(3+aR)\frac{(1+exp(-aR))}{aR}\right],\\
D_2=[exp(aR)+1]^2\left[\frac{1+exp(-aR)}{aR}\right]^2\left[3+aR-\frac{2aR}{1+exp(-aR)}\right],\nonumber
\label{E21}
\end{eqnarray}
and higher order terms are neglected. It is worth to note that the
above expansion is valid for low rotational energy states. Now, inserting the approximation expression (\ref{E20}) into equation (\ref{E18}) and changing the variables $r\rightarrow z$ through the mapping function $z=\exp{(-\nu x)}$, equation (\ref{E18}) turns to
\begin{eqnarray}
\frac{d^2R_{n\ell}(z)}{dz^2}+\frac{1}{z}\frac{dR_{n\ell}(z)}{dz}+\frac{1}{\left[\nu z(1+qz)\right]^2}\left\{\left(\frac{2\mu( q^2E_{n\ell}+qV_0)}{\hbar^2}-\frac{\ell(\ell+1)}{R^2}\left(q^2D_0-qD_1+D_2\right)\right)\right. z^2\nonumber\\
+\left(\frac{2\mu}{\hbar^2}(2qE_{n\ell}+V_0)-\frac{\ell(\ell+1)}{R^2}(2qD_0-D_1)\right)z+\left.\left(\frac{2\mu E_{n\ell}}{\hbar^2}-\frac{\ell(\ell+1)D_0}{R^2}\right)\right\}R_{n\ell}(z)=0.
\label{E22}
\end{eqnarray}
Before applying the AIM to this problem, we have to obtain the asymptotic wave functions and then transform equation (\ref{E22}) into a suitable form of the AIM. This can be achieved by the analysis of the asymptotic behaviours at the origin and at infinity. As a result the boundary conditions of the wave functions $R_{n\ell}(z)$ are taken as follows:
\begin{eqnarray}
R_{n\ell}(z)&\rightarrow& 0 \ \ \ \ \mbox{when}\ \ \ \ z\rightarrow -\frac{1}{q},\nonumber\\
R_{n\ell}(z)&\rightarrow& 0 \ \ \ \ \mbox{when}\ \ \ \ z\rightarrow 0,
\label{E23}
\end{eqnarray}
Thus, one can write the wave functions for this problem as
\begin{equation}
R_{n\ell}(z)=z^\alpha(1+qz)^\gamma F_{n\ell}(z),
\label{E24}
\end{equation}
where we have introduced parameters $\alpha$ and $\gamma$ defined by 
\begin{eqnarray}
\alpha&=&\frac{1}{\nu}\left[\frac{\ell(\ell+1)D_0}{R^2}-\frac{2\mu E_{n\ell}}{\hbar^2}\right]^{\frac{1}{2}},\nonumber\\
\gamma&=&\frac{1}{2}\left[1+\sqrt{1+\frac{4\ell(\ell+1)D_2}{\nu^2R^2q^2}}\right]^{\frac{1}{2}},
\label{E25}
\end{eqnarray}
for simplicity. By substituting equation (\ref{E24}) into equation (\ref{E22}), we have the second-order homogeneous differential equation of the form:
\begin{equation}
F_{n\ell}''(z)+\left[\frac{(2\alpha+1)+q(2\alpha+2\gamma+1)z}{z(1+qz)}\right]F_{n\ell}'(z)+\left[\frac{\frac{A}{q}+q(\alpha+\gamma)^2}{z(1+qz)}\right]F_{n\ell}(z)=0,
\label{E26}
\end{equation}
where $A=\frac{2\mu( q^2E_{n\ell}+qV_0)}{\hbar^2}-\frac{\ell(\ell+1)}{R^2}\left(q^2D_0-qD_1+D_2\right)$. Equation (\ref{E26}) is now suitable to an AIM solutions. By comparing this equation with equation (\ref{E1}), we can write the $\lambda_0(z)$ and $s_0(z)$ values and consequently; by means of equation (\ref{E7}), we may derive the $\lambda_k(z)$ and $s_k(z)$ as follows:
\begin{eqnarray}
\lambda_0(z)&=&-\left(\frac{(2\alpha+1)+qz(2\alpha+2\gamma+1)}{z(1+qz)}\right),\nonumber\\
s_0(z)&=&-\left(\frac{\frac{A}{q}+q(\alpha+\gamma)^2}{z(1+qz)}\right),\nonumber\\
\lambda_1(z)&=&-\left\{\frac{2\gamma+2\alpha+1}{z(1+qz)}+\frac{(\gamma+\alpha)^2q+A/q}{z(1+qz)}+\frac{2zq(\gamma+\alpha+1/2)-1-2\alpha}{z^2(1+qz)}\right.\nonumber\\
&&\left.q\frac{-2qz(\gamma+\alpha+1/2)-1-2\alpha}{z(1+qz)^2}-\frac{\left(2qz(\gamma+\alpha+0.5)+1+2\alpha\right)^2}{z^2(1+qz)^2}\right\}\nonumber\\
s_1(z)&=&\frac{\left(q(\gamma+\alpha)^2+A/q\right)\left(2qz(\gamma+\alpha+1/2)+1+2\alpha\right)}{z^2(1+qz)^2}-\frac{\left(q(\gamma+\alpha)^2+A/q\right)}{z^2(1+qz)^2}\\
\ldots etc.\nonumber
\label{E27}
\end{eqnarray}
The substitution of the above equations into equation (\ref{E10}), we obtain the first $\delta$ values as
\begin{equation}
\delta_o(z)=\left(\frac{\left(q^2(\alpha+\gamma)(4+\alpha+\gamma)+A+4q^2\right)\left(q^2(\alpha+\gamma)(2+\alpha+\gamma)+A+q^2\right)\left(q^2(\gamma+\alpha)^2+A\right)}{q^3z^3(1+qz)^3}\right)=0.
\label{E28}
\end{equation}
From the root of equation (\ref{E28}), we obtain the first relation between $\alpha$ and $\gamma$ as $\alpha_o=-\left[\gamma+\frac{1}{q}\sqrt{-A}\right]$. In a similar fashion, we can obtain other $\delta$ values and consequently establish a relationship between $\alpha_n$ and $\gamma$, $n=1, 2, 3,$ $\cdots$ as
\begin{eqnarray}
\delta_1(z)=
\left|
\begin{array}{lr}     
\lambda_2(z)&s_2(z) \\      
  \lambda_{1}(z)&s_{1}(z)
  \end{array}
  \right|=0\ \ \ \ \Rightarrow\ \ \ \ \alpha_1=-\left[\gamma+1+\frac{1}{q}\sqrt{-A}\right]\nonumber\\
   \delta_2(z)=
\left|
\begin{array}{lr}     
\lambda_3(z)&s_3(z) \\      
\lambda_{2}(z)&s_{2}(z)
\end{array}
\right|=0\ \ \ \ \Rightarrow\ \ \ \ \alpha_2=-\left[\gamma+2+\frac{1}{q}\sqrt{-A}\right]\nonumber\\
\ldots etc.
\label{E29}
\end{eqnarray}
The nth term of the above arithmetic progression is found to be
\begin{equation}
\alpha_n=-\left[\gamma+n+\frac{1}{q}\sqrt{-A}\right].
\label{E30}
\end{equation}
By substituting for $\alpha$ and $\gamma$, we obtain a more explicit expression for the eigenvalues energy as
\begin{equation}
E_{n\ell}=\frac{\hbar^2\ell(\ell+1)}{2\mu R^2}\left[D_0-\frac{D_1}{q}+\frac{D_2}{q^2}\right]-\frac{V_0}{q}-\frac{\hbar^2\nu^2}{8\mu}\left[\frac{\frac{1}{q\nu^2}\left[\frac{2\mu V_0}{\hbar^2}+\frac{\ell(\ell+1)}{R^2}(D_1-\frac{D_2}{q})\right]-{\chi(n,\ell)}^2}{\chi(n,\ell)}\right]^2,
\label{E31}
\end{equation}
where $\chi(n,\ell)=n+\frac{1}{2}+\frac{1}{2}\sqrt{1+\frac{4\ell(\ell+1)D_2}{q^2v^2R^2}}$. Let us now turn to the calculation of the wave functions. By comparing equation (\ref{E26}) with equation (\ref{E13}) we have the following:
\begin{equation}
t=-q\gamma,\ \ \ \ b=-q,\ \ \ \ N=-1,\ \ \ \ m=\alpha-\frac{1}{2},\ \ \ \ \sigma=2\alpha+1,\ \ \ \ \rho=2(\alpha+\gamma). 
\label{E32}
\end{equation}
Having determined these parameters, we can easily find the wave functions as
\begin{equation}
F_{n\ell}(z)=(-1)^nC_2\frac{\Gamma(2\alpha+n+1)}{\Gamma{(2\alpha+1)}}\ _2F_1\left(-n, 2(\alpha+\gamma)+n; 2\alpha+1; -qz\right),
\label{E33}
\end{equation}
where $\Gamma$ and $_2F_1$ are the Gamma function and hypergeometric function respectively. By using equations (\ref{E24}) and (\ref{E33}), the total radial wave function can be written as follows:
\begin{equation}
R_{n\ell}(r)=(-1)^nN_{n\ell}\frac{\left(1+qexp\left(\frac{R-r}{a}\right)\right)^\gamma}{exp\left(\alpha\frac{r-R}{a}\right)}\ {_2F_1\left(-n, 2(\alpha+\gamma)+n; 2\alpha+1; -qexp\left(\frac{R-r}{a}\right)\right)},
\label{E34}
\end{equation}
where $N_{n\ell}$ is the normalization constant. For a special case $\ell=0$, equation (\ref{E31}) reduces to 
\begin{equation}
E_n=-\frac{V_o}{q}-\frac{\hbar^2\nu^2}{8\mu}\left[\frac{2\mu V_o}{\hbar^2\nu^2q(n+1)}-(n+1)\right]^2=\frac{-\hbar^2}{2\mu}\left[\frac{V_o\mu}{q\hbar^2\nu(n+1)}+\frac{(n+1)\nu}{2}\right]^2,
\label{E35}
\end{equation}
and for the q-deformed Hulth$\acute{e}$n potential $\left(q\rightarrow-q; \nu\rightarrow\delta\right)$ we have
\begin{equation}
E_n=\frac{-\hbar^2}{2\mu}\left[\frac{V_o\mu}{q\hbar^2\delta(n+1)}-\frac{(n+1)\delta}{2}\right]^2,
\label{E36}
\end{equation}
which is identical to the ones obtained before using the factorization method \cite{BJ11}, SUSYQM approach \cite{NEW1, NEW2, NEW3}, quasi-linearization method \cite{NEW4} and Nikiforov-Uvarov method \cite{NEW5, NEW6}.
\begin{figure}[h!]
\centering
\includegraphics[height=66mm,width=105mm]{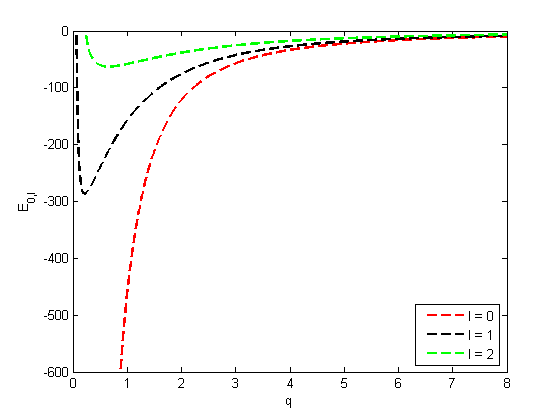}
\caption{\footnotesize The variation of the ground state $(n=0)$ energy level for various $\ell$ as a function of the deformation parameter. We choose $\mu=1fm^{-1}$, $V_0=40.5+0.13A fm^{-1}$, $A=40$, $R=1.25A^{1/3}$ and $a=0.65fm$. }
\end{figure}

\begin{figure}[h!]
\centering
\includegraphics[height=66mm,width=105mm]{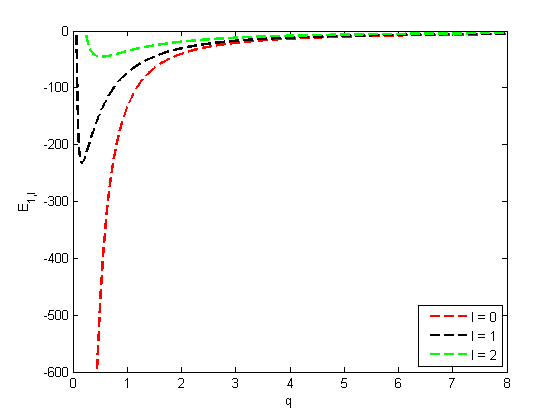}
\caption{\footnotesize The variation of the first excited $(n=1)$ energy state for various $\ell$ as a function of the deformation parameter.} 
\label{fig2}
\end{figure}

\begin{figure}[h!]
\centering
\includegraphics[height=66mm,width=105mm]{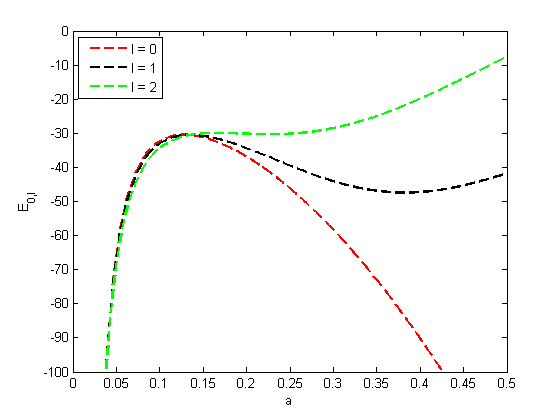}
\caption{\footnotesize The variation of the ground state energy state for various values of $\ell$ as a function of the diffuseness of the nuclear surface. We choose $\mu=1fm^{-1}$, $V_0=40.5+0.13A$ $fm^{-1}$, $A=40$, $R=1.25A^{1/3}$ and $q=1.5fm$.} 
\label{fig3}
\end{figure}
\begin{figure}[h!]
\centering
\includegraphics[height=66mm,width=105mm]{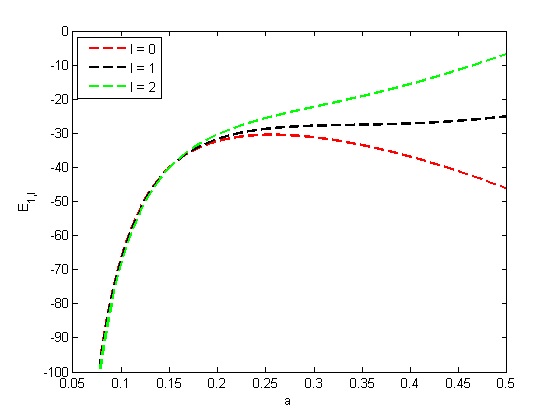}
\caption{\footnotesize The variation of the first excited energy state for various $\ell$ as a function of the diffuseness of the nuclear surface.} 
\label{fig4}
\end{figure}

\begin{figure}[h!]
\centering
\includegraphics[height=66mm,width=105mm]{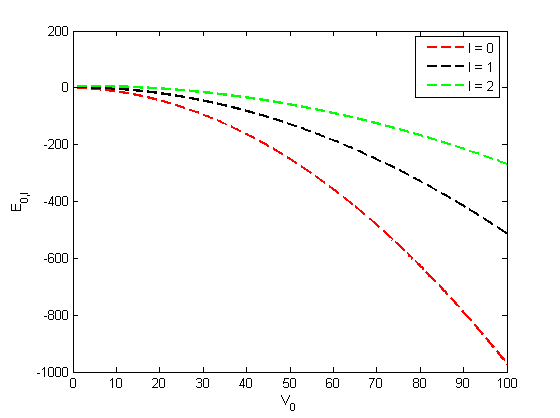}
\caption{\footnotesize The variation of the ground state energy level for various $\ell$ as a function of the potential depth $V_0$. We choose $\mu=1fm^{-1}$, $a=0.65fm$, $A=40$, $R=1.25A^{1/3}$ and $q=1.5fm$.} 
\end{figure}
\begin{figure}[h!]
\centering
\includegraphics[height=66mm,width=105mm]{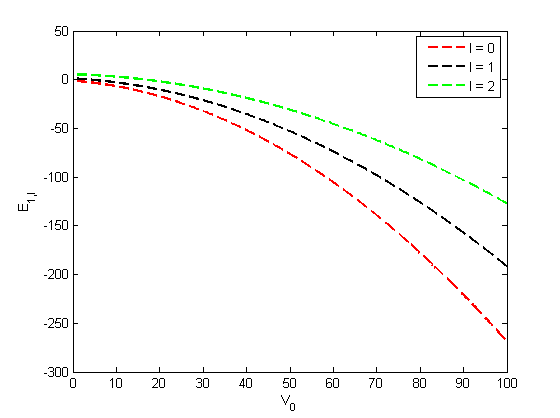}
\caption{\footnotesize The variation of the first excited energy state for various $\ell$ as a function of the potential depth $V_0$. }
\label{fig6}
\end{figure}

\begin{figure}[h!]
\centering
\includegraphics[height=66mm,width=105mm]{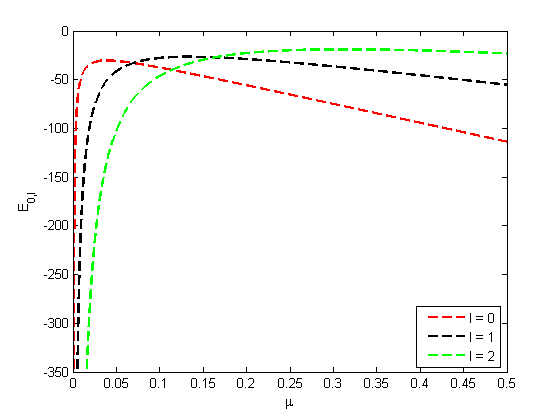}
\caption{\footnotesize The variation of the ground energy state as a function of the particle mass. We choose $a=0.65fm$, $V_o=40.5+0.13A fm^{-1}$, $A=40$, $R=1.25A^{1/3}$ and $q=1.5fm$. The radial quantum number is fixed to $n=0$.} 
\label{fig9}
\end{figure}

\begin{figure}[h!]
\centering
\includegraphics[height=66mm,width=105mm]{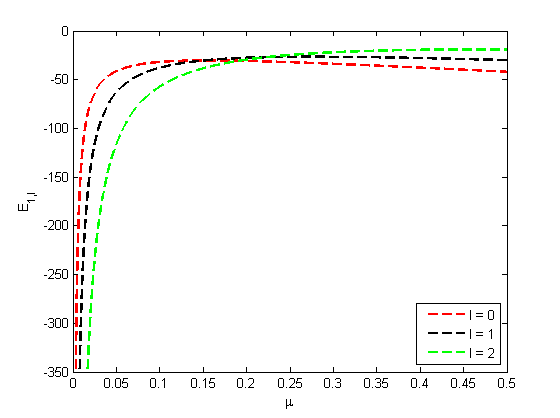}
\caption{\footnotesize The variation of the first excited energy spectrum as a function of the particle mass.} 
\label{fig10}
\end{figure}

To examine the behavior of the energy spectrum with the deformation parameter $q$, we plot the ground $n=0$ and first excited $n=1$ states for various values of $\ell$ and as  a function of deformation parameter $q$ as shown in Figures 1 and 2, respectively. In Figures 1 and 2, when the rotational quantum number $\ell$ and deformation parameter $q$ increases, the particle becomes less attractive or the energy is less negative (tends to continuum states). However, when $q<0.4$, the particle is strongly bound since the increasing $q$ shields the WS field. In Figures 3 and 4, we plot the ground and first excited states,  respectively for various $\ell$ and as function of $a$. We see that for the case $q=1.5$, the energy curve is strongly bound for wide range of $a$ when $\ell=0$ and then goes to less attractive. the depth of the attractive energy is much higher for small $\ell$. However, when $\ell=10$, the depth is small and the attractive energy has a short range of $a$ which then becomes purely repulsive. We conclude that the increament in the rotational quantum number $\ell$ leads to the attractive energy which is sensitive to the parameter $a$ and thus the selection of $a$ is  in the short range. In Figures 5 and 6, we plot energy versus $V_0$. We remark that for selected orbital state, the energy becomes more attractive with increasing $V_0$. In Figures 7 and 8, the energy is strongly attractive when the particle mass increases. As the rotational quantum number $\ell$ increases, the attractive energy speed up toward the positive energy (i.e., the energy becomes less attractive with the increasing $\ell$).  

We have seen that the approximate analytical bound states solutions of the $\ell-$wave Schr\"{o}dinger equation for the nuclear deformed Woods-Saxon potential can be solved by proper approximation to the centrifugal term within the framework of the AIM. Closed analytical forms for the energy eigenvalues are obtained and the corresponding wave functions have been presented in terms of hypergeometric functions. 

To test the accuracy of the present potential model, we present in Table 1 the approximate nonrelativistic energy states for various vibrational  $n=0,1,2,3$ and rotational $\ell$ states. As in nuclear physics, we used the parameter values $q=1$, $2\mu/\hbar^2=0.4727 MeV^{-1}fm^{-2}$, $R=1.285A_0^{1/3}$, $a=0.65$ and $V_0=40.5+0.13A_0$. For example, the energy states are calculated by means of the energy formula (31) using $V_0=45.7000$ and $V_0=47.7800$ for two nuclei of atomic masses $A_0=40$ and $A_0=56$, respectively. It is seen that when the vibrational quantum number $n$ increases then the energy increases toward the positive energy, i.e., it is becoming weakly bound (less negative). The entire bound states are calculated up to the continuum states where the number of negative bound states increases. The numerical results obtained in this work are found identical to the ones obtained before via Nikiforov-Uvarov method in Ref. $\cite{BJ14}$. Furthermore, the energy expression (31) provides the solution for the Hulthen potential if $q$ is being replaced by $-q$ $\cite{BJ21}$. We should also remark that the analytical energy formulae (31) and (58) in Ref. $\cite{BJ14}$ are identical and both provide identical numerical values in the MATLAB code. It is also found that the present approximation to the centrifugal term $1/r^2$ is valid only for the lowest energy states $\cite{BJ22}$. 

Finally, the method presented in this paper is an elegant and powerful technique. If there are analytically solvable potentials, it provides the closed forms for the eigenvalues and the corresponding eigenfunctions. However, the case if the solution is not available, the eigenvalues are obtained by using an iterative approach $\cite{BJ23, BJ24, BJ25}$.

We wish to thank the referees for their helpful suggestions and critics which have greatly helped us to improve the paper. One of the authors (BJF) acknowledges the efforts of Prof. Oyewumi K J, for his encouragements.

\begin{table}[!h]
\caption{The bound state energy eigenvalues $E_{n\ell}$ for some values of $n$ and $\ell$ with $R = 1.285 A_0^{1/3}$, $V_o=40.5+0.13A_0$, $a=0.65$, $2\mu/\hbar^2=0.4727MeV^{-1}fm^{-2}$ and $\hbar=6.5821\times 10^{-22}MeV.s$.}
\label{tab1}\vspace*{15pt} {
\begin{center}
\begin{tabular}{|c|c|c|c|c|c|c|c|}
\hline
{} &{} &{} &{} &{} & {} & {} & {}\\[-1.0ex] 
$n$ & $\ell$ & $E_{n\ell}$ $(A=40)$ & $E_{n\ell}$ $(A=56)$ & $n$ & $\ell$ & $E_{n\ell}$ $(A=40)$ &$E_{n\ell}$ $(A=56)$\\[2.5ex] \hline
0&0&-38.74580076&-41.69282644&2&0&-25.58756978&-26.37723930\\[1ex]
 &1&-21.08802055&-27.88702375&&1&-22.89764467&-22.97221589\\[1ex]
 &2&-6.800578910&-9.821051731&&2&-18.34331348&-16.99573336\\[1ex]
 &3&-0.179198496& 0.2326908620&&3&-12.94203150&-9.881694792\\[1ex]
 &4&&&&4&-7.576059522&-3.388054444\\[1ex]
 &5&&&&5&-2.951028070& 0.600789897\\[1ex]
 &6&&            &&6& 0.377741219&            \\[1ex] \hline  
1&0&-22.94937178&-24.07612820&3&0&-33.082779314&-33.75446842\\[1ex]
 &1&-17.82193874&-19.05824465&&1&-31.51835298&-30.95614994\\[1ex]
 &2&-11.08315873&-11.06363099&&2&-28.21109897&-25.78665449\\[1ex]
 &3&-5.012942659&-3.329348972&&3&-23.52936071&-19.02278925\\[1ex]
 &4&-0.753601214& 0.596581631&&4&-18.05628773&-11.68537314\\[1ex]
 &5& 1.057060553&            &&5&-12.37118321&-4.946616232\\[1ex]
 &6&&            &&6&-6.993259073&-0.620098829\\[1ex]
 &7&            &            &&7&-2.372969819& 1.676807238\\[1ex]
 &8&            &            &&8& 1.103404777&            \\[1ex]\hline
\end{tabular}
\end{center}
} \vspace*{-1pt}
\end{table}
\end{document}